# Subnanometric control of coupling between $WS_2$ monolayers with a molecular spacer

Sara A. Elrafei[1], Tom T. C. Sistermans[2,3], Alberto G. Curto[1,2,3*]

[1] Department of Applied Physics and Eindhoven Hendrik Casimir Institute, Eindhoven University of Technology, 5600 MB Eindhoven, The Netherlands

[2] Photonics Research Group, Ghent University-imec, Ghent, Belgium

[3] Center for Nano- and Biophotonics, Ghent University, Ghent, Belgium

* Corresponding author: A.G.Curto@TUe.nl

**Stacking monolayer semiconductors into heterostructures allows for control of their optical and electronic properties, offering advantages for nanoscale electronics, optoelectronics, and photonics. Specifically, adding a thin spacer between monolayers can yield bulk materials that retain interesting monolayer properties, such as a direct bandgap and a high emission quantum efficiency. The interaction mechanisms between monolayers, including interlayer coupling, charge transfer, and energy transfer, might be tuned through subnanometric control over the spacer thickness. Traditional spacer materials like bulk oxides or other layered materials can suffer from poor material interfaces or inhomogeneous thickness over large areas. Here, we use a spin-cast organic molecular spacer to adjust interlayer coupling in $WS_2$ monolayer stacks. We vary the molecular spacer thickness to tune the interlayer distance, significantly altering the optical properties of the resulting organic-inorganic heterostructures. Additionally, we demonstrate a dependence of the valence-band splitting on molecular spacer thickness manifested as a change in the energy difference between A and B excitons resulting from spin-orbit coupling and interlayer interactions. Our results illustrate the potential of molecular spacers to tailor the properties of monolayer heterostructures. This accessible approach opens new routes to advance atomically thin devices and could enable sensing technologies at the subnanometer scale.**

Heterostructures formed by stacking layers of two-dimensional materials like graphene, hexagonal boron nitride (hBN), and transition metal dichalcogenides (TMDs) are central to advancing nanoscale devices due to their exceptional properties and functionalities.[1–5] Heterostructures can exhibit enhanced performance exceeding that of each constituent material alone.[6–8] This versatility is particularly impactful in optoelectronics and sensing applications.[9–13] By controlling the interaction between layers, heterostructures can be designed so that the monolayers function independently or have strong interactions. The stacking configurations and the arrangement of the layers can be strategically chosen to tailor the electronic and optical properties. Different methods have been suggested to tune interlayer interactions, including electrical modulation, strain, twist angle, intercalation, and control over interlayer distance.[14–19] Controlling the interlayer distance provides a direct and effective way to modulate interlayer coupling. This approach offers precise tuning of the electronic and optical properties by adjusting the separation between layers, overcoming the challenges associated with other methods, such as complexity in fabrication and post-fabrication modifications.

Beyond the control of interlayer interactions, another important drive for heterostructure fabrication is preserving the interesting intrinsic properties of monolayer materials when transitioning to more complex and bulk layered systems.[20–22] Monolayer semiconductors such as $WS_2$ and $MoS_2$ have a direct bandgap, which becomes indirect with additional layers. Monolayers also show a high exciton binding energy attributed to their reduced dimensionality.[23–25] However, their atomically thin nature limits their optical absorption and emission, and their surface makes them sensitive to their environment and strain.[26–29] One strategy to circumvent these limitations is to stack monolayers into superlattices,[21,22] forming thicker films with densely packed layers while ensuring low interlayer coupling to preserve the unique properties of individual monolayers. Intercalating a spacer layer between two monolayers can prevent the transition to an indirect bandgap of few-layer and bulk TMD crystals. The dominant mechanism for interlayer interaction and its strength depend on interlayer distance[30] and can thus be controlled by modifying the spacer thickness. For example, at small interlayer distances below 1 nm, photoluminescence (PL) quenching can result from nonradiative processes such as Dexter-type energy transfer or charge transfer. Dexter transfer, in particular, arises from wavefunction overlap between adjacent $WS_2$ monolayers. For distances between 2 and 10 nm, Förster energy transfer by dipole-dipole interactions takes over. With increased separation, interlayer interactions become negligible and the monolayers can be regarded as uncoupled.[31–33]

Inorganic spacer materials like hBN,[30,34,35] $Al_2O_3$,[36] and graphene[37] are often used in heterostructures to adjust interlayer interactions. However, these materials pose challenges for precise thickness control and interface quality. For instance, van der Waals material spacers are limited to discrete thicknesses. Conventional dielectric spacers such as bulk oxides typically suffer from low material quality for thicknesses in the 1-nm range, even when deposited using atomic layer deposition. Another growing family of approaches relies on molecular spacers instead of atomic crystals or bulk materials. Using water adsorption to control van der Waals gaps presents a pioneering method for tailoring the properties of heterostructures.[38] However, processing could face limitations when heating or storing in vacuum are required. Organic molecular materials are alternative spacers providing a scalable method to tune the interlayer coupling precisely.[39,40] Molecular intercalation for chemical dedoping of TMD monolayers has been recently exploited to tune carrier density and interlayer coupling for improved functionality in bulk TMDs.[20] Despite such progress, obtaining monolayer stacks and thicker superlattices with desirable optical properties, such as a high exciton oscillator strength and quantum efficiency, remains challenging.

Here, we control the interlayer interaction between two $WS_2$ monolayers using tetracyanoquinodimethane (TCNQ) as a molecular spacer. We vary the molecular spacer thickness through spin coating at different molecular concentrations. We quantify PL quenching, report a redshift at low molecular concentrations, and use Raman spectroscopy to gain further insights into the interlayer interactions. Our analysis reveals how the molecular spacer influences valence-band splitting. We model the energy difference between the A and B excitons using spin-orbit and interlayer coupling terms to identify the effect of spacer thickness on the monolayer interactions. Integrating molecular spacers with monolayer semiconductors has therefore the potential to improve the functionality and performance of these materials and their heterostructures. Furthermore, hybrid organic-inorganic structures pave the way for sensing technologies leveraging two-dimensional materials to perform molecular-scale distance measurements through changes in their electronic, optical, and optoelectronic responses.

**TCNQ as a molecular spacer**

Our starting point is $WS_2$ monolayers exfoliated from bulk crystals onto a polydimethylsiloxane (PDMS) film and identified by wide-field fluorescence microscopy under illumination with a blue lamp. $WS_2$ monolayers show strong PL, which is notably weaker for bilayers, corresponding to the transition from a direct to an indirect band gap. We deposit TCNQ onto a monolayer via spin coating using methanol as the solvent (see Methods). TCNQ will serve later as a spacer between two monolayers (Figure 1a). Its planarity is crucial for its function as a spacer, allowing for homogeneous charge distribution and interaction across the layers.[41] To estimate the thickness of the molecular spacer, we performed atomic force microscopy (AFM) measurements of spin-cast TCNQ films on quartz substrates. The spacer thickness increases approximately linearly with TCNQ concentration (Figure 1c). TCNQ also provides *p*-type doping,[22] which enhances the emission quantum efficiency of monolayers without significantly altering the exciton energy and linewidth.[42–46] The controlled p-type doping introduced by TCNQ has been previously used for molecular engineering of $WS_2$ heterostructures for field-effect transistors.[41,47] First, we investigate the effect of molecular doping in individual monolayers before stacking. We compare the PL spectra of pristine and TCNQ-doped $WS_2$ monolayers (Supporting Section S1). The pristine monolayer exhibits emission attributed to neutral excitons ($X^0$, peak at 611.7 nm) and charged trions ($X^-$, peak at 614.4 nm), as the $WS_2$ crystal is originally an *n*-type semiconductor. After doping with TCNQ, the PL of a monolayer increases up to 2.5-fold due to a reduction of trion formation and a higher efficiency of neutral exciton emission (Supporting Section S1).[42,46] As the TCNQ concentration increases, we observe a PL enhancement that saturates at higher concentrations (Figure 1b).

To tune interlayer coupling, we create stacks of two monolayers with varying TCNQ spacer concentrations by transferring another $WS_2$ monolayer onto a doped monolayer (Methods). The final assembly is encapsulated in PDMS. We then compare the photoluminescence from monolayers and bilayers with different spacer thicknesses (Figure 1d): assembled bilayers with TCNQ concentrations of 2 and 0.1 mM, an assembled bilayer without TCNQ, and a directly exfoliated bilayer with a natural van der Waals gap. The PL spectra of the single monolayer and the TCNQ-spaced bilayer with a concentration of 2 mM are similar, indicating that the large distance between the two monolayers leads to almost no interlayer interaction. For a TCNQ concentration of 0.1 mM, the PL spectrum reveals a pronounced exciton redshift due to the lower spacer thickness and stronger coupling. Without TCNQ, the bilayer shows a slightly more redshifted spectrum, which resembles the direct transition observed in natural bilayers more closely. The indirect transition centered at 700 nm that dominates the PL spectrum for natural bilayers is not visible in artificially stacked monolayers. After this initial observation of coupling in bilayers with molecular spacers, we analyze next the impact of interlayer interactions on the optical properties in more detail.

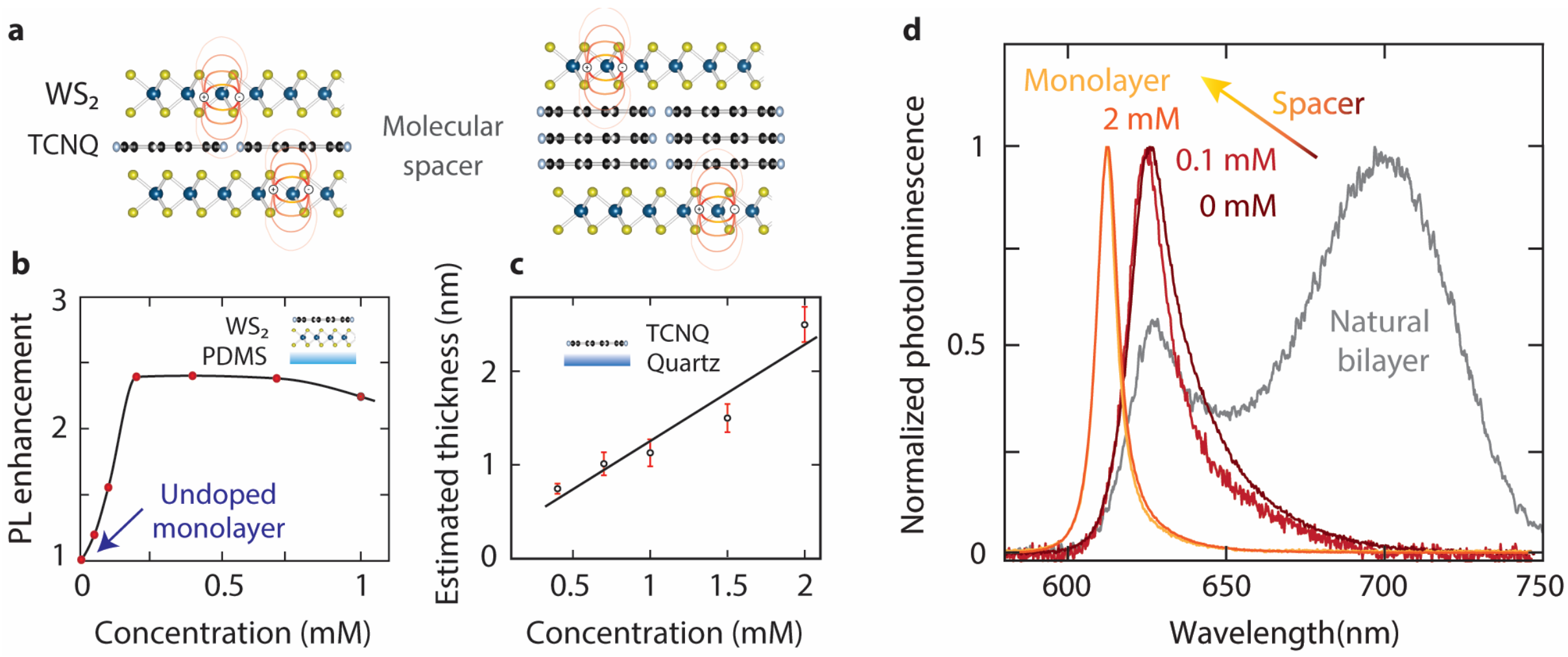

**Figure 1 | Stacked $WS_2$ monolayers with a molecular spacer. a,** Two monolayers separated by a TCNQ molecular spacer of varying thickness. Shorter interlayer distances result in stronger interlayer interaction due to overlapping wavefunctions. **b,** PL enhancement with increasing TCNQ doping of single $WS_2$ monolayers calculated as the ratio of the peak intensities after/before doping for each monolayer. **c,** Spacer thickness estimated using atomic force microscopy of spin-cast TCNQ on amorphous quartz. Error bars indicate standard deviation within an AFM scan. **d,** Normalized PL spectra for decreasing spacer thickness showing an exciton energy shift due to stronger interlayer interaction: a single monolayer (yellow), assembled monolayers with molecular spacer concentrations

of 2 mM (orange) and 0.1 mM (red), and no molecular spacer (dark red) compared to a natural bilayer (gray).

**Optical signatures of interlayer coupling: spectral shift and quenching**

As an indication of interlayer interactions, we first exploit Raman spectroscopy. The Raman scattering spectrum of $WS_2$ features in-plane ($E_{2g}$) and out-of-plane ($A_{1g}$) vibrational modes; the energy difference between both peaks serves as a sensitive indicator of the number of layers and interlayer interactions for TMDs.[48–50] For monolayer $WS_2$, the $E_{2g}$ and $A_{1g}$ modes appear at 354.17 $cm^{-1}$ and 417.18 $cm^{-1}$, respectively (Figure 2a, yellow).[51,52] While the $E_{2g}$ mode remains unaffected by the number of layers, the $A_{1g}$ mode exhibits a clear blueshift upon transitioning to a bilayer (Figure 2a, black), indicating lattice stiffening due to the introduction of the second layer. The observed separations between the Raman peaks in the monolayer (63.01 $cm^{-1}$) and bilayer (64.77 $cm^{-1}$) align well with reported values.[53,54] Next, we investigate how the TCNQ spacer concentration between $WS_2$ monolayers influences the Raman peak separation (Figure 2a, orange and red). The Raman peaks move apart as the TCNQ concentration decreases, reflecting a stronger interlayer coupling. At a concentration of 1 mM, the Raman peak separation is 63.42 $cm^{-1}$, similar to the monolayer situation and indicating weak interaction. At a concentration of 0.2 mM, the Raman peak separation increases to 64.74 $cm^{-1}$, approaching the value for a natural bilayer. These results prove that the molecular spacer concentration affects interlayer coupling, thereby altering the optical properties of $WS_2$ as shown next.

We focus on the change of PL intensity and spectrum with varying molecular spacer thicknesses. The PL spectra for bilayers with low TCNQ spacer concentrations exhibit substantial quenching in PL intensity and a spectral shift (Figure 2b). At 2 mM, the PL intensity of the bilayer is comparable to that of two uncoupled doped monolayers, suggesting minimal interlayer interaction and independent doping of both monolayers. At low concentrations, however, the reduced spacer thickness brings the $WS_2$ layers into closer proximity, enhancing interlayer interactions and reducing PL intensity. Quenching can be quantified by the ratio of the peak intensity of a doped monolayer to that of the corresponding artificial bilayer. This quenching ratio reaches approximately 15 for a spacer concentration of 0.1 mM (Figure 2c, right axis). As the spacer thickens, the reduction in interlayer coupling leads to a lower quenching ratio. Concurrently, we observe a blueshift of the A-exciton peak as the spacer thickens (Figure 2c, left axis). In summary, the presence of both quenching and wavelength shift underscores the sensitivity of $WS_2$ optical properties to the molecular spacer, which directly controls interlayer distance and coupling.

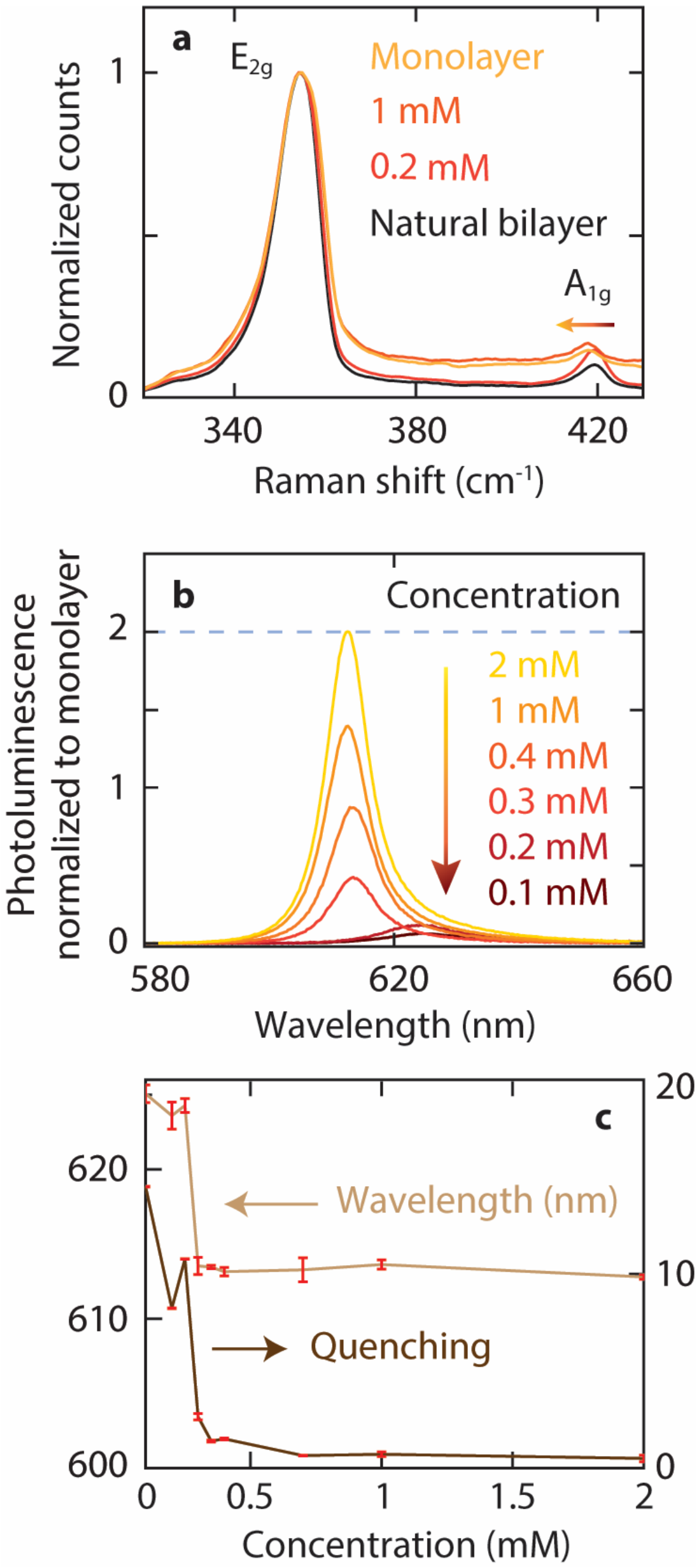

**Figure 2 | Photoluminescence quenching and shift for stacked $WS_2$ monolayers with a molecular spacer. a,** Raman shifts for different spacer configurations: an individual monolayer, a natural $WS_2$ bilayer, and artificial $WS_2$ bilayers with TCNQ spacer concentrations of 0.2 and 1 mM. **b,** PL spectra of artificial $WS_2$ bilayers normalized to the corresponding doped monolayer for different TCNQ

concentrations. The dashed line indicates the PL peak intensity of two uncoupled but doped monolayers; quenching occurs for values below this line. **c,** Quenching and wavelength shift as a function of molecular concentration. Right: quenching calculated as the ratio of the peak intensity in the doped monolayer to that of the artificial bilayer. Left: A-exciton spectral peak position.

To achieve a more detailed and statistically significant understanding of interlayer coupling, we analyze spatially resolved hyperspectral PL images. By recording a spectrum at every point within a specified sample region, we obtain statistics of several exciton properties within that area. We examine different TCNQ concentrations, focusing on the bilayer areas enclosed by dashed blue lines in Figure 3a. The PL intensity maps illustrate the evolution from strongly coupled layers with quenching at 0.1 mM to loosely coupled layers at 2 mM, which emit nearly twice the intensity of a monolayer. Using hyperspectral analysis, we identify correlations between different properties for varying spacer thicknesses. We find that the PL intensity and the peak energy are correlated (Figure 3b, left): at lower spacer concentrations, there is a decrease in PL intensity as the peak energy decreases. Similarly, there is a correlation between intensity and linewidth (Figure 3b, right): the exciton emission broadens as it dims due to increased phonon interactions, greater disorder within the layers, and enhanced non-radiative rates.

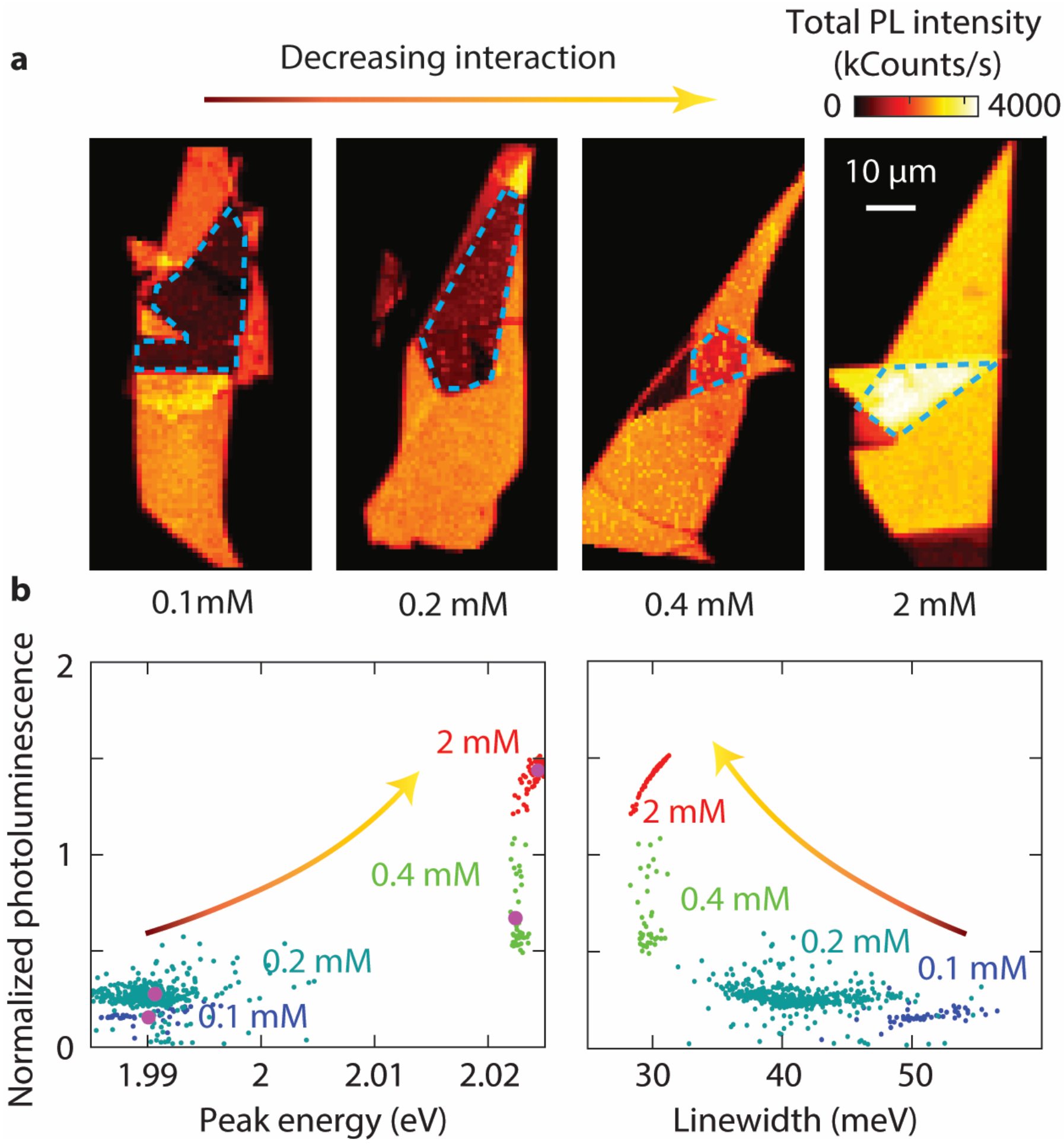

**Figure 3 | Hyperspectral imaging of monolayer stacks with increasing molecular spacers**. **a,** Spectrally integrated PL maps. Dashed blue lines denote stacked monolayer areas. **b,** Scatter plots derived from hyperspectral PL images of $WS_2$ bilayers with different spacer thicknesses. Left: PL intensity versus peak energy. Right: PL intensity versus linewidth. The normalized PL is defined as the ratio of the maximum intensity in the bilayer to that of the corresponding doped monolayer.

**Transmission and valence-band splitting**

We shift focus to transmission spectroscopy to provide complementary information to PL on how molecular spacers affect the optical properties of stacked monolayers. The transmittance contrast at the A-exciton dip is defined as the difference between the minimum transmittance at the A exciton and the baseline at longer wavelengths. As the TCNQ concentration increases in the stacked $WS_2$ monolayers (Figure 4a), we observe a higher transmittance contrast accompanied by a reduction in the A-exciton linewidth. By fitting the spectra with four Lorentzian peaks (Supporting Section S2), we quantify the increase in transmittance contrast at the A-exciton peak, which indicates a weakening in interlayer coupling (Figure 4b). As in the case of PL, there is also a blueshift of the transmission dip with increasing concentration (Figure 4a). These changes suggest significant implications for the electronic structure.

The band diagram of monolayer TMDs features valence-band splitting at the K point arising from spin-orbit coupling (SOC). The energy difference between A and B excitonic transitions (Figure 5a) results from such SOC splitting ($\Delta E_{\mathrm{SOC}}$). Interacting monolayers have an additional contribution to valence-band splitting ($\Delta E_{\mathrm{LC}}$) from interlayer coupling (LC).[55,56] Interlayer distance thus governs LC without significantly affecting SOC, as SOC arises from the interaction between electron spin and orbital motion. We model the energy difference between the A and B excitons in coupled monolayers as $\Delta E = E_{\mathrm{B}} - E_{\mathrm{A}} = \sqrt{(\Delta E_{\mathrm{SOC}})^2 + (\Delta E_{\mathrm{LC}})^2}$.[56,57] The LC strength, related to the interlayer distance *d*, is described by $\Delta E_{\mathrm{LC}} = E_{\mathrm{LC}_0}\, \mathrm{e}^{-d/\tau_{\mathrm{LC}}}$, where *d* is proportional to the molecular concentration. We write then $\Delta E = \sqrt{(\Delta E_{\mathrm{SOC}})^2 + (E_{\mathrm{LC}_0}\, \mathrm{e}^{-d/\tau_{\mathrm{LC}}})^2}$, where $\Delta E_{\mathrm{SOC}}$ represents the constant contribution from SOC and $E_{\mathrm{LC}_0}$ and $\tau_{\mathrm{LC}}$ are the characteristic energy and distance constants of LC.

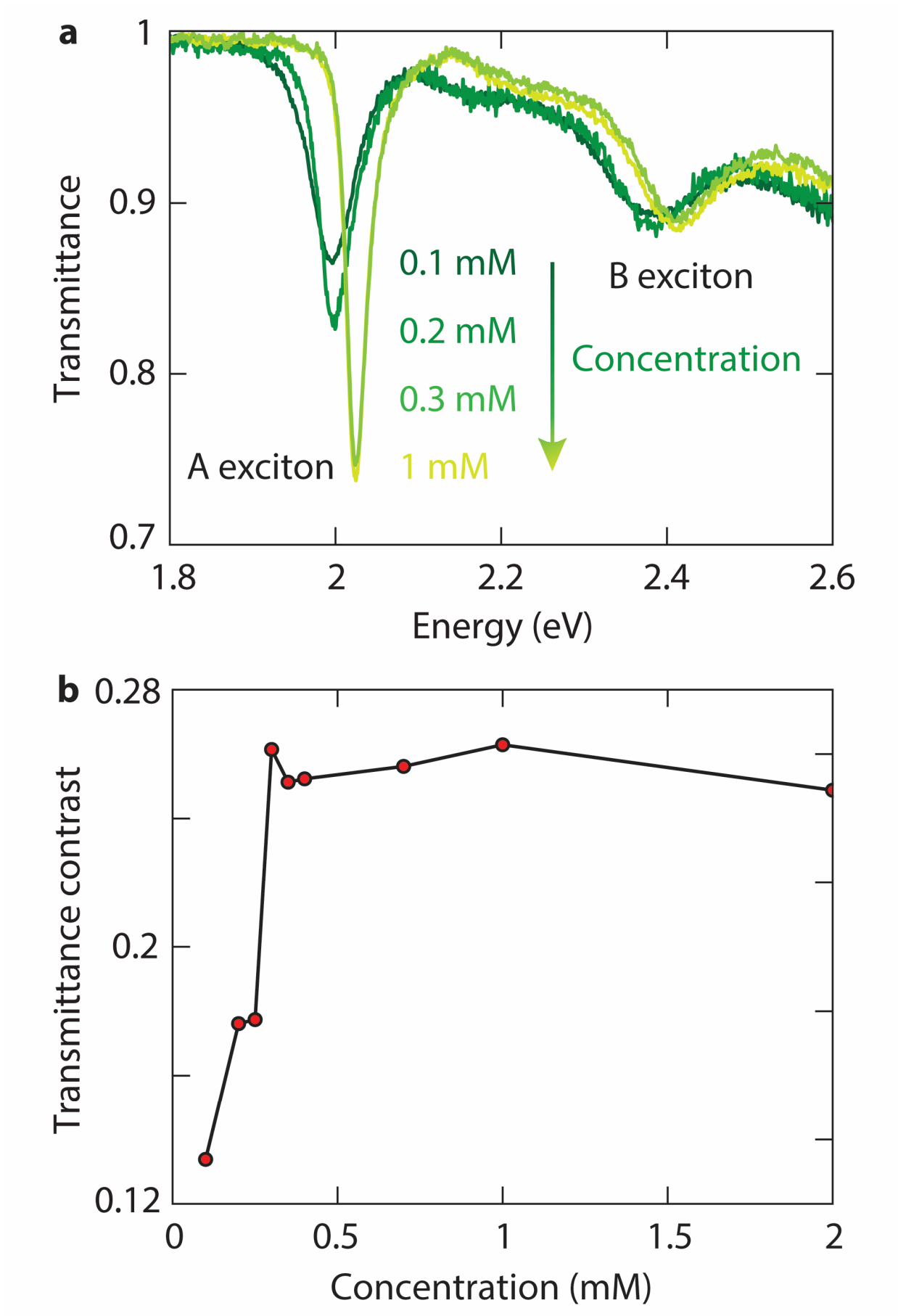

**Figure 4 | Transmission changes with molecular spacer concentration in stacked $WS_2$ monolayers.** **a,** Transmittance spectra for stacked monolayers with different TCNQ concentrations produced by spin coating. **b,** Transmittance contrast obtained by fitting the experimental spectra with four Lorentzian peaks. The A-exciton peak shows lower light absorption for coupled monolayers at low concentrations.

We investigate how the molecular spacer influences the valence-band splitting by retrieving the energy difference between the A and B excitons in our experimental transmittance spectra and fitting it to the model above (Figure 5b). The fitted curve (black line) includes a constant spin-orbit coupling (SOC, orange area) and an exponential dependence on interlayer coupling (LC, pink area). We extract $E_{\mathrm{SOC}}$ = 382.0 ± 1.5 meV, which is in line with other experimental and theoretical findings.[58] Furthermore, the fitting parameters are $E_{\mathrm{LC}_0}$ = 170.0 ± 249.2 meV and $\tau_{\mathrm{LC}}$ = 0.10 ± 0.06 mM. At sufficiently high molecular spacer concentrations, the band splitting is dominated by SOC alone (orange line), closely resembling that of a $WS_2$ monolayer, where the splitting is approximately 383.5 meV. As SOC originates from intrinsic atomic interactions within each monolayer, it remains unaffected by interlayer separation. At low molecular concentrations, the contribution from LC becomes visible, although the total band splitting remains below the natural bilayer value (dashed gray line). In this regime, the reduced spacing enhances interlayer orbital overlap and hybridization, resulting in an additional splitting contribution beyond SOC.

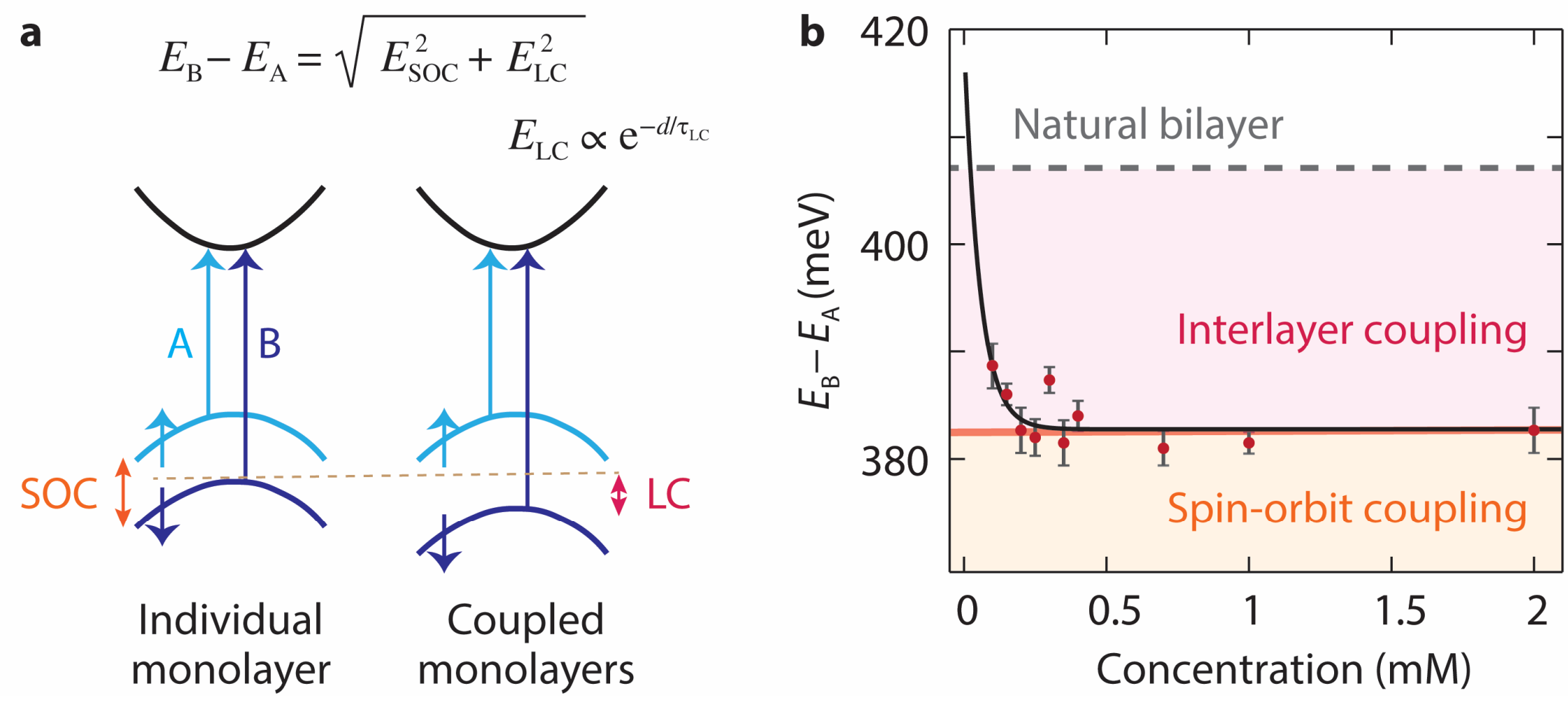

**Figure 5 | Band splitting dependence on the molecular spacer due to interlayer coupling**. **a,** Schematic band diagram for a $WS_2$ monolayer and two coupled monolayers around the K point showing splitting of the valence band due to spin-orbit (SOC) and interlayer (LC) couplings. **b,** Energy difference between the A and B excitons as a function of molecular spacer concentration for stacked monolayers. Black line: fit including constant intralayer spin-orbit coupling (SOC, orange area) and interlayer coupling with an exponential dependence on distance (LC, pink area). Band splitting for a natural bilayer shown for reference (dashed gray lines). Error bars represent the standard deviation across different samples and take into account the spectrometer resolution.

## Conclusion

We have successfully manipulated the interlayer coupling in heterostructures consisting of $WS_2$ monolayers and organic molecules as a subnanometric spacer. This alternative approach to traditional spacer materials like hBN and $Al_2O_3$ significantly changes the optical properties of the stacked monolayers. Additionally, the use of TCNQ as a molecular spacer facilitates charge transfer providing *p*-type doping. We have demonstrated control over photoluminescence in these heterostructures by modifying its peak energy, linewidth, and emission quantum efficiency. We observed a spectral redshift with increasing molecular spacer thickness. Our solution-based method enables manipulation of interlayer interaction by adjusting the molecular concentration, with the layers transitioning from strongly coupled at lower concentrations to loosely coupled at higher concentrations. We also reported a considerable impact on the valence-band splitting, as evidenced by the variations in the energy difference between A and B excitons, which we described using a model incorporating both spin-orbit and interlayer coupling. While this approach shows promise, certain limitations remain, particularly in terms of achieving consistent homogeneity and ensuring precision and repeatability during fabrication. These challenges highlight areas where further processing optimization is needed to realize its full potential. More broadly, engineering spacers in heterostructures with molecular materials offers a powerful tool for fine-tuning semiconductor properties. The demonstrated sensitivity to interlayer distance opens opportunities for molecular-scale sensors and subnanometric rulers with optical, electronic, or optoelectronic readout.

**Methods**

**Sample preparation.** We exfoliate a bulk $WS_2$ crystal with *n*-type doping (HQ Graphene) into a monolayer using tape (Nitto Denko, SPV 9205) and deposit it on an optically transparent PDMS film (Gel-Pak, PF-80-X4) placed on a glass slide. As the molecular spacer, we use *p*-type dopant molecules, 7,7,8,8-tetracyanoquinodimethane (TCNQ, Ossila Ltd). We employ various concentrations of TCNQ in methanol (Merck). For example, for the preparation of a 2-mM solution, we dissolve 8.2 mg of TCNQ powder in 20 mL of methanol. Subsequently, 30 µL of this solution are deposited onto a $WS_2$ monolayer on PDMS lying on a glass slide, followed by spinning at 500 RPM for 1 minute. This low speed ensures the formation of a thin, uniform film while minimizing any potential damage to the TMD layer, as confirmed by atomic force microscopy (Supporting Section S3). To construct monolayer stacks including the molecular spacer, we use the all-dry viscoelastic stamping method[59–61] to transfer the top monolayer onto the bottom monolayer covered with TCNQ using an optical microscope equipped with two *xyz* micrometric stages for precise placement. We leave the top PDMS film on top of the stack, thus fully encapsulating it in PDMS, which prevents direct exposure to the environment. Finally, the structure is heated to 70 °C on a hotplate for homogeneous contact between the layers**.**

**Optical measurements**. We use a home-built confocal microscope for photoluminescence, transmission spectroscopy, and hyperspectral imaging. For photoluminescence excitation, we utilize a continuous-wave laser at 532 nm (Cobolt Samba). Using neutral density filters, the power reaching the sample is in the 1−100 μW range depending on PL efficiency changes due to doping and quenching. The excitation laser is cleaned using a band-pass filter (Thorlabs, FLH532-4), reflected toward the sample by a beam splitter (Chroma, 21014 silver non-polarizing 50/50 bs), and focused onto the sample using an objective with adjustable cover-glass correction (Nikon, 40x CFI Plan Fluor ELWD, NA = 0.6). Photoluminescence is filtered from the excitation laser with a long-pass filter (Thorlabs, FELH0550) and collected in epifluorescence configuration. For transmission measurements, we illuminate the sample from the bottom with a white light source using Köhler illumination through an objective with adjustable cover-glass correction (Nikon, 20x CFI Plan Fluor ELWD, NA = 0.45). The transmitted light is then collected through the top objective and coupled into an optical fiber with a core size of 50 μm serving as the confocal pinhole. This fiber is connected either to a spectrometer (Andor Shamrock 303i spectrograph with a 300 lines/mm grating and an Andor Newton 970 EMCCD camera) or an avalanche photodiode (Micro Photon Devices, PDM50). Raman spectroscopy relied on a 1800 lines/mm grating instead.

**Supporting Information**

The Supporting Information is available free of charge at (…): Doping of monolayer $WS_2$; Permittivity and transmission spectrum of monolayer $WS_2$; Atomic force microscopy.

**Acknowledgments**

This work was financially supported by the Netherlands Organization for Scientific Research (NWO) through an NWO START-UP grant (740.018.009) and the European Research Council (ERC) under the European Union's Horizon 2020 Research and Innovation Program (Grant Agreement 948804, CHANSON). We thank Bingying You and Rasmus H. Godiksen for stimulating discussions.

**Conflict of Interest**

The authors declare no conflict of interest.

**Data Availability Statement**

The data that support the findings of this study are available from the authors upon reasonable request.